\documentclass[showpacs,aps,pra,amsfonts,amsmath,amssymb,superscriptaddress,floatfix]
{revtex4}
\usepackage[dvips]{graphicx}

\newcommand{\affA}{%
     Quantum Information Technology Group, 
     National Institute of Information and Communications Technology
     (NICT), \\
     4-2-1 Nukui-kitamachi, Koganei, Tokyo 184-8795, Japan}

\begin{document}

\title{EPR beams and photon number detector:
Toward synthesizing optical nonlinearity}

\author{M. Sasak}
\address{\affA}%
\author{K. Wakui}
\address{\affA}%
\author{J. Mizuno}
\address{\affA}%
\author{M. Fujiwara}
\address{\affA}%
\author{M. Akiba}
\address{\affA}%

\begin{abstract}
We present the two kinds of experimental results. 
One is a continuous variable dense coding experiment, 
and 
the other is a photon number detector with high 
linearity response, 
the so called charge integration photon detector (CIPD).  
They can be combined together to be a potential tool 
for implementing the cubic phase gate 
which is an important gate element 
to synthesize the measurement induced nonlinearity 
for photonic quantum information processing. 
\end{abstract}

\maketitle


\section{Introduction}

  One of the important tasks in quantum optics and quantum 
information precessing is to implement any desired optical 
nonlinearity. 
Currently available nonlinear media, however, suffer from 
weak strengths and high losses.  
It was recently shown that 
nonlinear interaction between photons can be implemented 
using linear optical elements, ancilla photons, and 
post selection based on the output of single-photon detectors, 
now referred to as the measurement induced nonlinearity.  
\cite{KLM01}. 
For example, 
a nondeterministic Kerr interaction can be effected 
in this way. 
Although this nondeterministic nonlinear interaction works only 
a fraction of total events, 
such operations can be performed off-line 
from the quantum computation.

\begin{figure}\label{CubicPhaseGate}
  \includegraphics[height=.3\textheight]{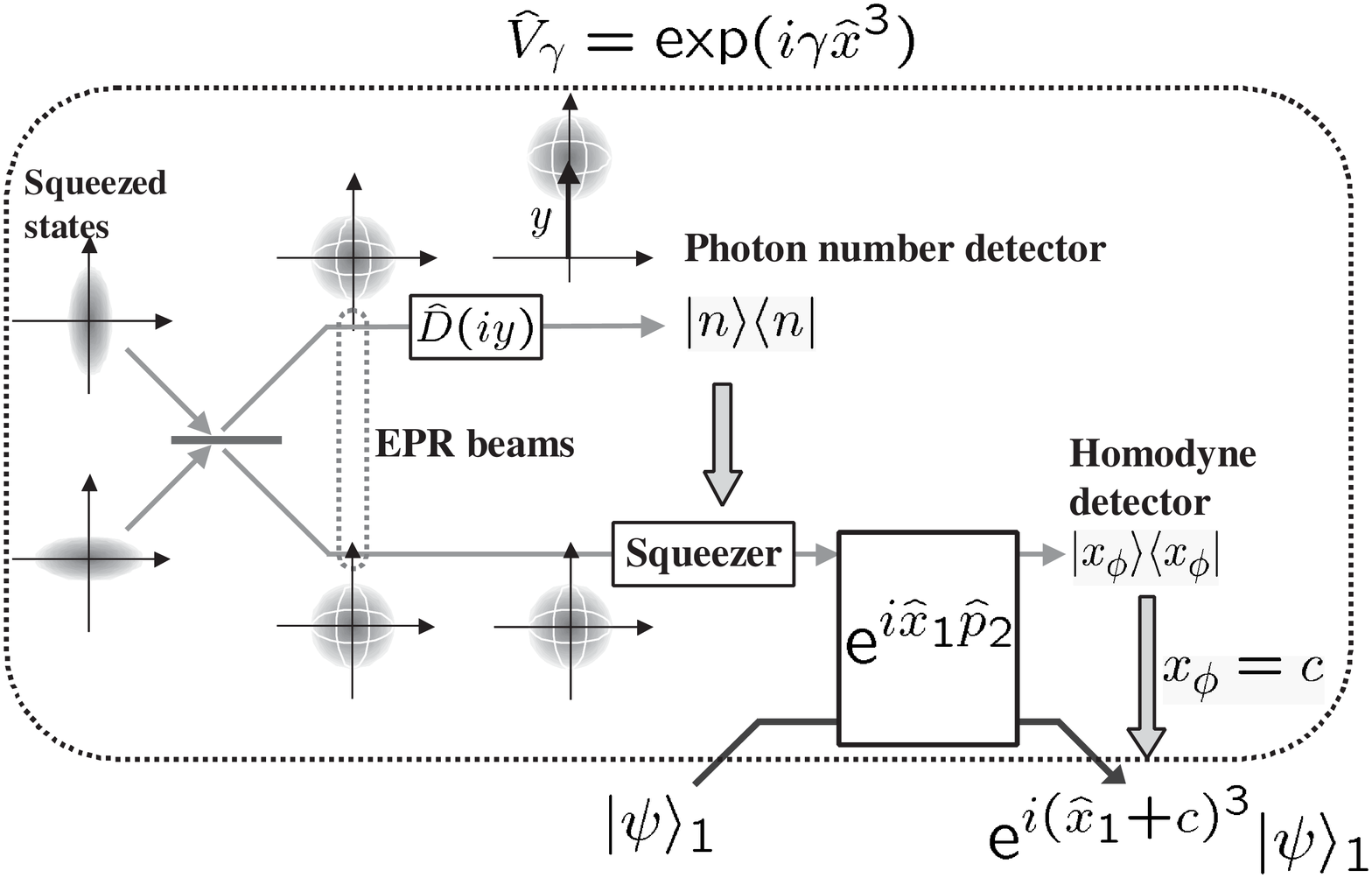}
  \caption{Schematic diagram of the cubic phase gate based on 
the measurement induced nonlinearity.}
\end{figure}

This original idea has been extended recently for the 
continuous variables of photonic field 
\cite{Gottesman01,BartlettSanders02}. 
In \cite{Gottesman01}, 
it was shown the cubic phase gate 
$\hat V_\gamma={\rm exp}( i \gamma\hat x^3)$ 
where $\hat x$ is the quadrature amplitude, 
can be implemented 
using the two mode squeezed state (EPR beams), 
the interactions specified by 
the Hamiltonians up to the bilinear order of 
the creation and anihilation operators 
$\{\hat a, \hat a^\dagger\}$, 
and 
the measurements 
by a photon number detector and a homodyne detector 
(see Fig. \ref{CubicPhaseGate}). 
It is also known that 
once a nonlinear Hamiltonian with a cubic or higher polynomial 
of $\{\hat a, \hat a^\dagger\}$ is obtained, 
it is possible to build up arbitrary multi-mode Hamiltonians 
by cascading this with 
the bilinear Hamiltonians 
and the beam splitters 
\cite{LloydBraunstein99}. 
The bilinear Hamiltonians 
are more or less in a range of current technologies. 
A big challenge is to synthesize a nonlinear Hamiltonian 
such as the cubic phase gate. 
This gate opens the possibilities 
of implementing quantum operations 
not only on the photonic states in the discrete spectrum 
but also on the states in the continuous spectrum. 
The latter includes Gaussian input states 
which are essential signal carriers in optical communications.

We present our recent work on the two key elements 
for the cubit phase gate:  
the continuous variable dense coding experiment with the EPR beams 
and a photon number detector.

\section{Continuous variable dense coding}

Figure \ref{CubicPhaseGate} shows a circuit to implement 
the cubic phase gate.  
One of the EPR beams first undergoes the displacement operation 
described by the operator $\hat D(\alpha)$. 
One of the encoded EPR beams is then detected 
by a photon number detector 
with the precision satisfying 
$\Delta n \ll n^{1/3}$. 
The measurement result of $n$ photons 
projects the other beam into a certain non-Gaussian state.  
This state is further transformed via a single mode squeezer, 
and is then coupled with a target input state 
$\vert\psi\rangle_1$ through a unitary process of 
a bilinear interaction $\mathrm{e}^{i\hat x_1 \hat p_2}$. 
One of the output mode is finally measured by a homodyne detection. 
According to the measurement result, 
the target state $\vert\psi\rangle_1$ can be modulated by 
the cubic phase shift.

\begin{figure}\label{DenseCodingCircuit}
  \includegraphics[height=.27\textheight]{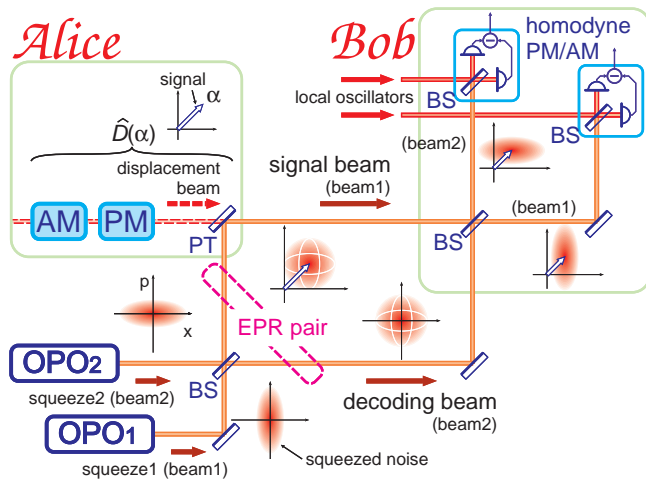}
  \caption{Schematic diagram of dense coding experiment using
squeezed vacuum states.}
\end{figure}

It is still difficult to perform this whole operation. 
We first work on the preparation of the displaced EPR beams, 
and the characterization of them, 
which is nothing but the continuous variable dense coding scheme 
\cite{BraunsteinKimble00}. 
Its schematic is shown in Fig. \ref{DenseCodingCircuit}.  
Two independent squeezed vacua 
from the optical parametric oscillator (OPO) cavities 
are combined to create EPR beams. 
For the OPO, a 10\,mm long potassium niobate (KNbO$_3$) crystal 
is employed. 
A light source is a Ti:sapphire laser (Coherent, MBR--110)
whose fundamental wavelength is 860\,nm. 
The OPO cavities are pumped by the second harmonic beam 
generated by a frequency doubler (Coherent, MBD--200).
The two squeezed beams are generated at fundamental wavelength. 
One of the EPR beams is super-imposed 
via a high reflectivity mirror ($T_\mathrm{PT}\approx 1\,\%$) 
with a bright beam with the amplitude and phase modulations. 
This setup realizes the displacement operation on 
the signal beam. 
The encoded EPR beams are decoded by the Bell measurement unit 
where the encoded EPR beams are first separated into 
two squeezed states via a 50\,:\,50 beamsplitter, 
and are then detected 
by two homodyne detectors
to extract the signal in amplitude- and phase-quadratures.

\begin{figure}\label{DenseCodingExpData}
  \includegraphics[height=.27\textheight]{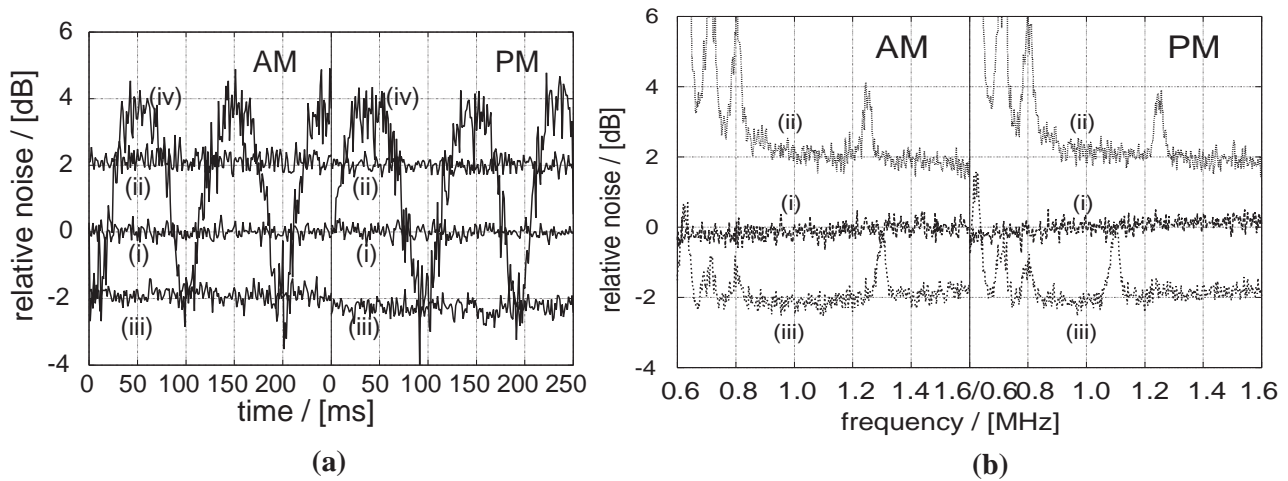}
  \caption{
(a) Outputs from the Bell measurement in the time domain;
(i)\,shot noise, 
(ii)\,EPR noise, 
(iii)\,squeezed beam with LO locked,
and 
(iv)\,squeezed beam. 
(b) Outputs from the Bell measurement in the frequency domain;
(i)\,shot noise, 
(ii)\,EPR noise, 
and 
(iii)\,squeezed beam with LO locked.
The vertical scale is normalized by the shot noise level
around 1.1\,MHz.
The peaks at 1.25\,MHz and in low frequency region
(${<}\,0.8\,\mathrm{MHz}$) are due to
the technical noises of the laser rather than the quantum noise.}
\end{figure}

The experimental results are shown 
in Fig. \ref{DenseCodingExpData}. 
Figure \ref{DenseCodingExpData}(a) 
shows the noise power of the two orthogonal quadratures 
around 1.1\,MHz in the time domain. 
As seen, 
the EPR noise (ii) is larger than
the shot noise (i) and also is phase independent.
This EPR noise can be canceled in the Bell measurement, 
resulting intwo separable squeezed states with 
about 2dB squeezing, (iii) and (iv), 
below the shot noise level. 
Figure \ref{DenseCodingExpData}(b) 
shows the noise power spectra in the frequency domain. 
The AM and PM signals (iii) 
at 1.3\,MHz and 1.1\,MHz, respectively, 
buried in the vacuum noise (i) 
are decoded by the Bell measurement. 
Thus we have successfully performed encoding and decoding 
assisted by the entanglement in the two orthogonal quadratures.

\section{Charge integration photon detector}

Now we move to the other important element, 
a photon number detector. 
Many approaches have been taken 
to construct photon number resolving detectors.  
For the visible light region, 
there is a nice device called VLPC 
which has a very high quantum efficiency, 
and could directly observe nonclassical light statistics
\cite{Waks_etal04}.  
There are also several other candidates 
based on sophisticated mechanisms, 
such as quantum dot array FET 
\cite{Shields_etal00}, 
and the superconducting edge sensor 
\cite{Miller_etal03}. 
Recently another practical scheme has been demonstrated, 
which is based on cascading commercially available Si-APDs 
\cite{Achilles_etal03,Fitch_etal03}. 
All these devices have both merits and demerits 
depending on our purposes.

Here we would like to add another candidate to this list. 
Our approach is very straightforward. 
We convert photons into electrons with low noise 
using a photodetector at cryogenic temperature 
either Si-APD in the linear mode 
for the visible and near infrared window 
or 
InGaAs p-i-n photo-diode for the telecom-fiber window. 
We then amplify photo-electrons 
by a very low noise charge integration amplifier.  
We call this scheme the charge integration photon detector 
(CIPD).

\begin{figure}\label{SiAPDCircuit_ProdDistr}
  \includegraphics[height=.3\textheight]{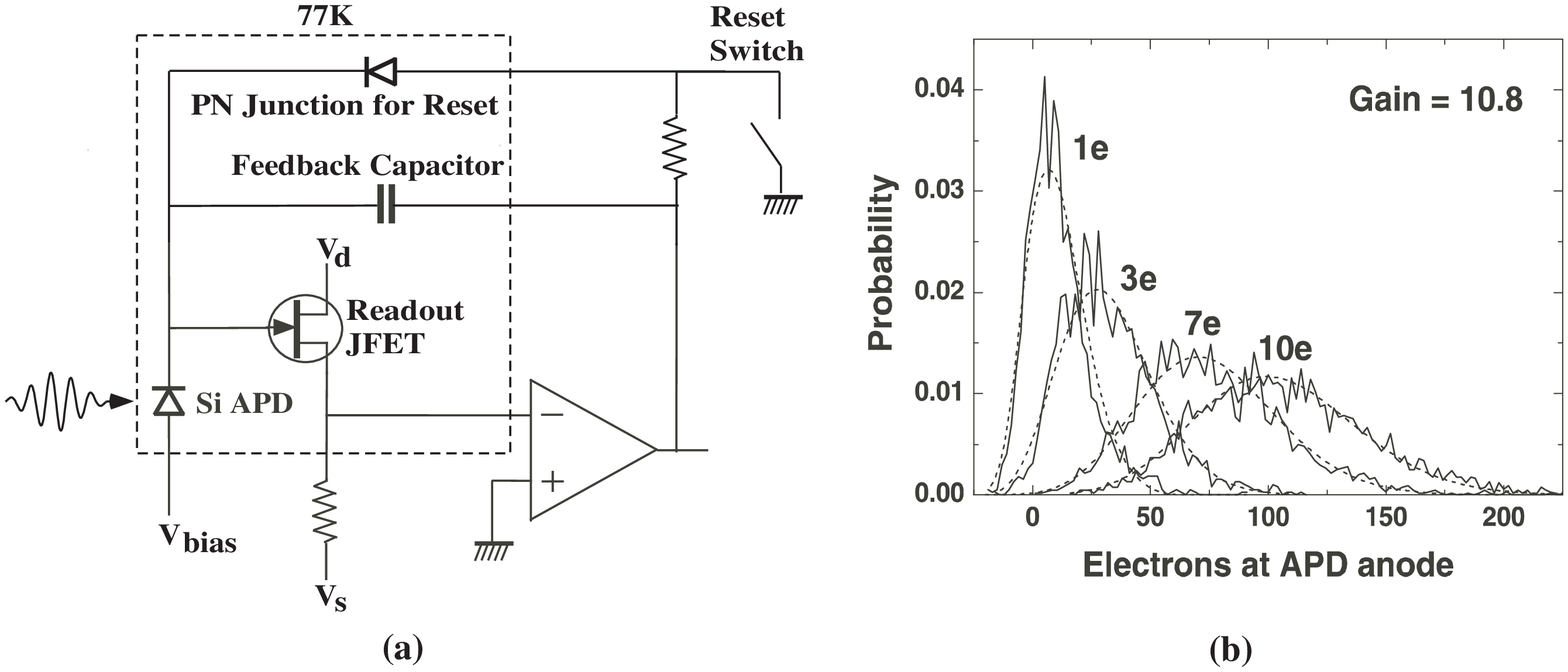}
  \caption{
   (a) Schematic diagram of CIPD. 
   (b) Probability distributions of photoelectron.}
\end{figure}

In Fig. \ref{SiAPDCircuit_ProdDistr}(a), 
we show a schematic diagram of CIPD in the case 
of Si-APD for the visible and near infrared window. 
The readout circuit is based on a capacitive transimpedance 
amplifier, 
where 
we use a Si junction field effect transister (JFET) 
as preamplifier, 
and a diode and a relay switch for reset. 
We removed as many of the materials and devices 
that generate leak current, stray capacitance, and 
dielectric polarization noises as possible from the circuit. 
As a result the readout circuit 
has ultralow current-noise at low frequencies 
and a low input capacitance of 1\,pF.  
So the Si-APD can be operated at very low gains. 
We use it with the gain about 10, 
corresponding to the bias voltage of 90\,V. 
The Si-APD used is a 200\,$\mu$m-diameter bare chip made by 
Matsusada Precision Inc. 
The quantum efficiency is estimated to be about 60\% 
at 635\,nm wavelength, 
which was indirectly determined 
from the room-temperature quantum efficiency (69\%) 
given in the manufacturer's catalog. 
That is, 
the drop of the output current at 77\,K was compared with 
the one at room temperature. 

Figure \ref{SiAPDCircuit_ProdDistr}(b) 
shows the photon number probability distributions 
for four kind of pulse intensities.  
(Precisely speaking, 
this corresponds to the number of 
input referred photoelectron. 
To convert them to the photon number probability distributions, 
we have to estimate the quantum efficiency more precisely.) 
These data were obtained 
by integrating the photoelectrons created by each pulse of 
2,000 light pulse events  
with repetition of 20Hz. 
The light source is a 635-nm light-emitting diode. 
Unfortunately, we could not have succeeded precise discrimination 
of photon number yet. 
At present we could have achieved very low noise characteristics, 
and high-linearity response to the number of incident photons.  
An upper-limit dark current of 1 e/s was obtained 
from the maximum variation of the output voltage in 1sec 
at 77K.

In order to distinguish photon number precisely,  
we have to reduce the readout circuit noise, 
currently 7 electrons at 20Hz sampling rate, 
further by 1/3,  
so as to improve the S/N ratio greater than 4. 
This is possible by reducing the total stray capacitance of 
the system.

\section{Conclusion}

We have presented two basic elements, 
the continuous variable dense coding experiment, 
and the charge integration photon detector (CIPD).  
They can be combined together to be a powerful tool 
for implementing the cubic phase gate 
which is an important gate element 
for photonic quantum information processing.


The authors acknowledge the financial support by 
the Ministry of Public Management, Home Affairs, Posts and 
Telecommunications, 
and 
the CREST project of Japan Science and Technology.  



\bibliography{sample}

\IfFileExists{\jobname.bbl}{}
 {\typeout{}
  \typeout{******************************************}
  \typeout{** Please run "bibtex \jobname" to optain}
  \typeout{** the bibliography and then re-run LaTeX}
  \typeout{** twice to fix the references!}
  \typeout{******************************************}
  \typeout{}
 }

\end{document}